\documentclass{article}
\usepackage{spconf,amsmath,amssymb,graphicx}
\usepackage{color,cite,balance,url,booktabs,siunitx}

\usepackage{hyperref}

\DeclareMathOperator*{\argmin}{argmin}

\title{End-to-End Multi-Speaker Speech Recognition with Transformer}
\name{Xuankai Chang$^1$, Wangyou Zhang$^{2}$, Yanmin Qian$^{2}$, Jonathan Le Roux$^{3}$, Shinji Watanabe$^{1}$ \thanks{Wangyou Zhang and Yanmin Qian were supported by the China NSFC project No.U1736202.}}
\address{
    $^1$Center for Language and Speech Processing, Johns Hopkins University, USA\\
    $^2$MoE Key Lab of Artificial Intelligence \& \\SpeechLab, Department of Computer Science and Engineering, Shanghai Jiao Tong University, China\\
    $^3$Mitsubishi Electric Research Laboratories (MERL), USA\\
    {\small\texttt{xchang14@jhu.edu, \{wyz-97, yanminqian\}@sjtu.edu.cn, leroux@merl.com, shinjiw@jhu.edu}}
}
\begin{document}
\ninept
\maketitle
\begin{abstract}

Recently, fully recurrent neural network (RNN) based end-to-end models have been proven to be effective for multi-speaker speech recognition in both the single-channel and multi-channel scenarios. In this work, we explore the use of Transformer models for these tasks by focusing on two aspects. First, we replace the RNN-based encoder-decoder in the speech recognition model with a Transformer architecture. Second, in order to use the Transformer in the masking network of the neural beamformer in the multi-channel case, we modify the self-attention component to be restricted to a segment rather than the whole sequence in order to reduce computation. Besides the model architecture improvements, we also incorporate an external dereverberation preprocessing, the weighted prediction error (WPE), enabling our model to handle reverberated signals. Experiments on the spatialized wsj1-2mix corpus show that the Transformer-based models achieve $40.9\%$ and $25.6\%$ relative WER reduction, down to $12.1\%$ and $6.4\%$ WER, under the anechoic condition in single-channel and multi-channel tasks, respectively, while in the reverberant case, our methods achieve $41.5\%$ and $13.8\%$ relative WER reduction, down to $\num[round-mode=places,round-precision=1]{16.50}\%$ and $\num[round-mode=places,round-precision=1]{15.24}\%$ WER.

\end{abstract}
\section{Introduction}
\label{sec:intro}

Deep learning techniques have dramatically improved the performance of separation and automatic speech recognition (ASR) tasks related to the cocktail party problem \cite{Experiment-Cherry1953}, where the speech from multiple speakers overlaps.
Two main scenarios are typically considered, single-channel and multi-channel. In single-channel speech separation, various methods have been proposed, among which deep clustering (DPCL) based methods \cite{DeepClustering-Hershey2015} and permutation invariant training (PIT) based methods \cite{Permutation-Yu2017} are the dominant ones. For ASR, methods combining separation with single-speaker ASR as well as methods skipping the explicit separation step and building directly a multi-speaker speech recognition system have been proposed, using either the hybrid ASR framework \cite{Recognizing-Yu2017,Monaural-Chang2018,Analysis-Menne2019} or the end-to-end ASR framework \cite{End-Settle2018,End2End-Seki2018,End-Chang2019}. In the multi-channel condition, the spatial information derived from the inter-channel differences can help distinguish between speech sources from different directions, which makes the problem easier to solve. Several methods have been proposed for multi-channel speech separation, including DPCL-based methods using integrated beamforming \cite{Tight-Drude2017} or inter-channel spatial features \cite{Multichannel-Wang2018}, and a PIT-based method using a multi-speaker mask-based beamformer \cite{Recognizing-Yoshioka2018}. For multi-channel multi-speaker speech recognition, an end-to-end system was proposed in \cite{MIMO-Chang2019}, called MIMO-Speech because of the multi-channel input (MI) and multi-speaker output (MO). This system consists of a mask-based neural beamformer frontend, which explicitly separates the multi-speaker speech via beamforming, and an end-to-end speech recognition model backend based on the joint CTC/attention-based encoder-decoder \cite{Joint-Kim2017} to recognize the separated speech streams.
This end-to-end architecture is optimized via only the connectionist temporal classification (CTC) and cross-entropy (CE) losses in the backend ASR, but is nonetheless able to learn to develop relatively good separation abilities.

Recently, Transformer models \cite{Attention-Vaswani2017} have shown impressive performance in many tasks, such as pretrained language models \cite{Improving-Radford2018,Bert-Devlin2018}, end-to-end speech recognition \cite{Improving-Nakatani2019,Comparative-Karita2019}, and speaker diarization \cite{Endtoend-Fujita2019}, surpassing the long short-term memory recurrent neural networks (LSTM-RNNs) based models. One of the key components in the Transformer model is self-attention, which computes the contribution information of the whole input sequence and maps the sequence into a vector at every time step. Even though the Transformer model is powerful, it is usually not computationally practical when the sequence length is very long. It also needs adaptation for specific tasks, such as the subsampling operation in encoder-decoder based end-to-end speech recognition. However, for signal-level processing tasks such as speech separation and enhancement, subsampling is usually not a good option, because these tasks need to maintain the original time resolution.

In this paper, we explore the use of Transformer models for end-to-end multi-speaker speech recognition in both the single-channel and multi-channel scenarios. First, we replace the LSTMs in the encoder-decoder network of the speech recognition module with Transformers for both scenarios. Second, in order to also apply Transformers in the masking network of the neural beamforming module in the multi-channel case, we modify the self-attention layers to reduce their memory consumption in a time-restricted (or local) manner, as used in \cite{Effective-Luong2015,Time-Povey2018,Monaural-Chang2018}. To the best of our knowledge, this work is the first attempt to use the Transformer model for tasks such as speech enhancement/separation with such very long sequences. Another contribution of this paper is to improve the robustness of our model in reverberant environments. To do so, we incorporate an external dereverberation method, the weighed prediction error (WPE) \cite{Generalization-Yoshioka2012}, to preprocess the reverberated speech. The experiments show that this straightforward method can lead to a performance boost for reverberant speech.

\begin{figure*}[htb]
\vspace{-0.6cm}
\begin{minipage}[b]{0.49\linewidth}
  \centering
  \label{fig:model1}
  \centerline{\includegraphics[width=0.9\linewidth]{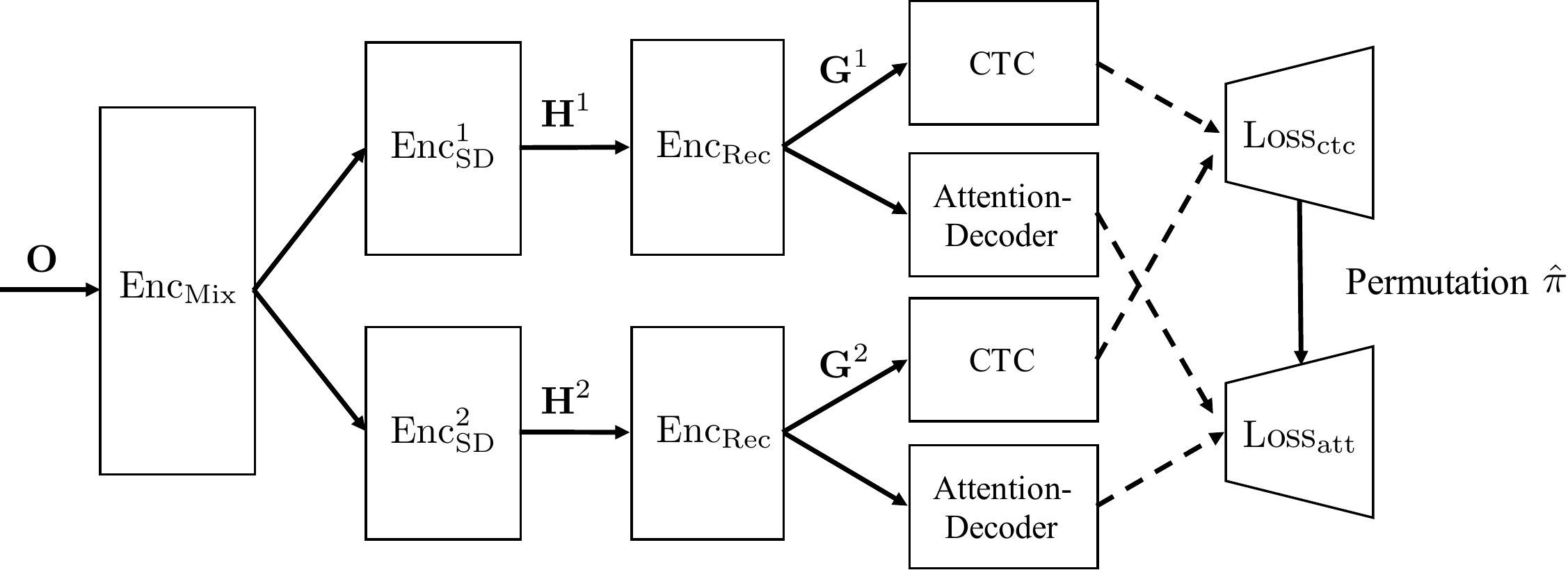}}
  \vspace{-0.15cm}
  \caption{End-to-end single-channel multi-speaker model in the 2-speaker case. The speaker-differentiating encoder (Enc$_{\text{SD}}$), recognition encoder (Enc$_{\text{Rec}}$), and decoder are either RNNs or Transformers.}
  \label{fig:scmc-model}
\end{minipage}
\hspace{0.02\linewidth}
\begin{minipage}[b]{0.49\linewidth}
  \centering
  \label{fig:model2}
  \centerline{\includegraphics[width=\linewidth]{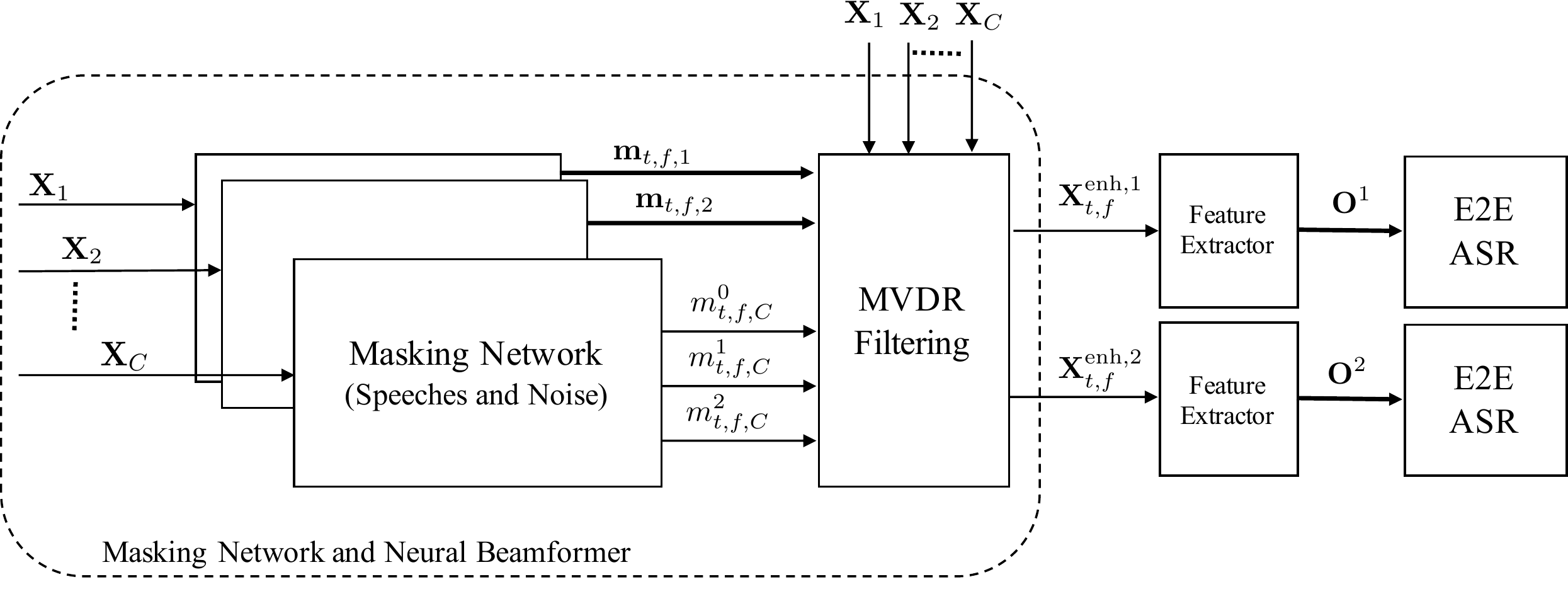}}
  \vspace{-0.15cm}
  \caption{End-to-end multi-channel multi-speaker model in the 2-speaker case. The masking network and end-to-end ASR network are based on either RNNs or Transformers.}
  \label{fig:mimo-model}
\end{minipage}
\vspace{-0.6cm}
\end{figure*}

\section{End-to-End Multi-speaker ASR}
\label{sec:model-arch}

In this section, we review the end-to-end speech recognition models for both the single-channel \cite{End2End-Seki2018,End-Chang2019} and multi-channel \cite{MIMO-Chang2019} tasks. For both tasks, we denote by $J$ the number of speakers in the input speech mixture.

\subsection{Single-channel Multi-speaker ASR}
\label{ssec:monaural-speech}

In this subsection, we briefly introduce the end-to-end single-channel multi-speaker speech recognition model proposed in \cite{End2End-Seki2018,End-Chang2019}, shown in Fig.\ref{fig:scmc-model}. The model is an extension of the joint CTC/attention-based encoder-decoder framework \cite{Joint-Kim2017} to recognize multi-speaker speech. The input $\mathbf{O}=\{\mathbf{o}_1,\dots,\mathbf{o}_T\}$ is the single-channel mixed speech feature. In the encoder, the input feature is separated and encoded as hidden states $\mathbf{G}^{j}, j=\{1,\dots,J\}$ for each speaker. The computation of the encoder can be divided into three submodules:
\begin{align}
    \mathbf{H} &= \text{Encoder}_{\text{Mix}}(\mathbf{O}), \\
    \mathbf{H}^{j} &= \text{Encoder}_{\text{SD}}^{j}(\mathbf{H}),\, j=1,\dots,J,\\
    \mathbf{G}^{j} &= \text{Encoder}_{\text{Rec}}(\mathbf{H}^{j}),\, j=1,\dots,J.
\end{align}
Encoder$_{\text{Mix}}$ first maps the input $\mathbf{O}$ to some high dimensional representation $\mathbf{H}$. Then $J$ speaker-differentiating encoders Encoder$_{\text{SD}}^{j}$ extract each speaker's speech $\mathbf{H}^{j}$. Finally, Encoder$_{\text{Rec}}$ transforms each $\mathbf{H}^{j}$ into the embeddings $\mathbf{G}^{j}=\{\mathbf{g}^{j}_1,\dots,\mathbf{g}^{j}_L\}$, $L \leq T$ with subsampling. The attention-based decoder then takes these hidden representations to generate the corresponding output token sequences $\mathbf{Y}^{j}=\{y^{j}_1,\dots,y^{j}_N\}$. For each embedding sequence $\mathbf{G}^{j}$, the recognition process is formalized as follows:
\begin{align}
    \mathbf{c}^{j}_n &= \text{Attention} (\mathbf{e}^{j}_{n-1}, \mathbf{G}^{j}) ,\\
    \mathbf{e}^{j}_{n} &= \text{Update}(\mathbf{e}^{j}_{n-1}, \mathbf{c}^{j}_{n-1}, y^{j}_{n-1}), \\
    y^{j}_n &\sim \text{Decoder} (\mathbf{e}^{j}_n, y^{j}_{n-1}),
\end{align}
in which $\mathbf{c}^{j}_n$ denotes the context vector and $\mathbf{e}^{j}_{n}$ is the hidden state of the decoder at step $n$. To determine the permutation of the reference sequences $\mathbf{R}^{j}$, permutation invariant training (PIT) is performed on the CTC loss right after the encoder \cite{End2End-Seki2018,End-Chang2019}:
\begin{align}
    \hat{\pi} = \argmin_{\pi \in \mathcal P} \sum_{j} \mathrm{Loss}_{\text{ctc}} (\mathbf{Z}^j, \mathbf{R}^{\pi(j)}),\; j=1,\dots,J, \label{eq:pit-ctc}
\end{align}
where $\mathbf{Z}^{j}$ is the sequence obtained from $\mathbf{G}^{j}$ by linear transform to compute the label posterior distribution, $\mathcal{P}$ is the set of all permutations on $\{1,\dots,J\}$, and $\pi(i)$ is the $i$-th element of permutation~$\pi$.
The model is optimized with both CTC and cross-entropy losses:
\footnotesize \begin{align}
    \mathcal{L} &= \sum_j \left( \lambda \mathrm{Loss}_{\text{ctc}} (\mathbf{Z}^j, \mathbf{R}^{\hat{\pi}(j)}) + (1-\lambda) \mathrm{Loss}_{\text{att}} (\mathbf{Y}^j, \mathbf{R}^{\hat{\pi}(j)}) \right), \label{eq:loss} 
\end{align} \normalsize
where $0 \leq \lambda \leq 1$ is an interpolation factor, and $\mathrm{Loss}_{\text{att}}$ is the cross-entropy loss of the attention-decoder.

\subsection{Multi-channel Multi-speaker ASR}
\label{ssec:mimo-speech}

In this subsection, we review the model architecture of the MIMO-Speech end-to-end multi-channel multi-speaker speech recognition system \cite{MIMO-Chang2019}, shown in Fig.\ref{fig:mimo-model}. The model takes as input the microphone-array signals from an arbitrary number $C$ of sensors. The model can be roughly divided into two modules, namely the frontend and the backend. The frontend is a mask-based multi-source neural beamformer.  For simplicity of notation, we denote the noise as the $0$-th source in the mixture signals. First, the monaural masking network estimates the masks $\mathbf{M}_c^j$ for every source $j\!=\!0,1,\dots,J$ on each channel $c\!=\!1,\dots,C$ from the complex STFT of the multi-channel mixture speech, $\mathbf{X}_c = (x_{t,f,c})_{t,f} \in \mathbb{C}^{T \times F}$, where $1\! \leq\! t \!\leq\! T$ and $1 \!\leq\! f \!\leq\! F$ represent the time and frequency indices, as follows:
\begin{align}
    \mathbf{M}_c = \text{MaskNet}(\mathbf{X}_c),
\end{align}
where $\mathbf{M}_c = (m^j_{t,f,c})_{t,f,j} \in [0,1]^{T \times F \times (J+1)}$. Second, the multi-source neural beamformer separates each source from the mixture based on the MVDR formalization \cite{Optimal-Souden2009}. The estimated masks of each source are used to compute the corresponding power spectral density (PSD) matrices $\mathbf{\Phi}^{j}$ for $j\! \in\! \{0, \dots, J\}$ \cite{NTT-Yoshioka2015,Neural-Heymann2016,Improved-Erdogan2016}:
\begin{align}
        \mathbf{\Phi}^{j} (f) &= \frac{1}{\sum_{t=1}^T m^{j}_{t,f}} \sum_{t=1}^T m^{j}_{t,f} \mathbf{x}_{t,f} \mathbf{x}^{H}_{t,f} \; \in \mathbb{C}^{C \times C},
\end{align}
where $\mathbf{x}_{t,f} = (x_{t,f,c})_{c}\in\mathbb{C}^C$, $m^{j}_{t,f} = \frac{1}{C} \sum_{c=1}^C m^{j}_{t,f,c}$ and $^H$ represents the conjugate transpose. The time-invariant filter coefficients $\mathbf{g}^{j}(f)$ for each speaker $j$ are then computed from the PSD matrices:
\begin{align}
        \mathbf{g}^j(f) &= \frac{(\sum _{i \neq j} \mathbf{\Phi}^{i} (f))^{-1} \mathbf{\Phi}^{j}(f)}{\text{Tr}((\sum _{i \neq j} \mathbf{\Phi}^{i} (f))^{-1} \mathbf{\Phi}^{j}(f))} \mathbf{u} \; \in \mathbb{C}^C, \label{eq:mvdr-filter}
\end{align}
where $1 \leq j \leq J$, and $\mathbf{u} \in \mathbb{R}^C$ is a vector representing the reference microphone derived from an attention mechanism \cite{Multichannel-Ochiai2017}. The beamforming filters $\mathbf{g}^{j}$ can be used to obtain the enhanced signal $\hat{s}^{j}_{t,f}$ for speaker $j$, which is further processed to get 
the log mel-filterbank with global mean and variance normalization ($\text{LMF}(\cdot)$):
\begin{align}
    \hat{s}^{j}_{t,f} &= (\mathbf{g}^{j}(f))^{H} \mathbf{x}_{t,f} \; \in \mathbb{C}, \\
    \mathbf{O}^{j} &= \text{LMF}(|\hat{\mathbf{S}}^{j}|)),
\end{align}
where $\hat{\mathbf{S}}^{j}$ is the short-time Fourier transform (STFT) of $\hat{s}^{j}$.

The backend ASR module maps the speech feature $\mathbf{O}^{j}=\{\mathbf{o}^{j}_1,\dots,\mathbf{o}^{j}_T\}$ of each speaker $j$ into the output token sequences $\mathbf{Y}^{j}=\{y^{j}_1,\dots,y^{j}_N\}$. The computation of the speech recognition is very similar to the process for the single-channel case described in Sec.~\ref{ssec:monaural-speech}, except that the encoder is a single path network and does not have to separate the input feature using Encoder$_{\text{SD}}$.

Similar to the single-channel model, the permutation order of the reference sequences $\mathbf{R}^{j}$ is determined by (\ref{eq:pit-ctc}). The whole MIMO-Speech model is optimized only with ASR loss as in (\ref{eq:loss}).

\section{Transformer with Time-restricted Self-Attention}
\label{sec:self-attention}

In this section, we describe one of the key components in the Transformer architecture, the multi-head self-attention \cite{Attention-Vaswani2017}, and the time-restricted modification \cite{Time-Povey2018} for its application in the masking network of the frontend.

Transformers employ the dot-product self-attention for mapping a variable-length input sequence to another sequence of the same length, making them different from RNNs. The input consists of queries $Q$, keys $K$, and values $V$ of dimension $d^{\text{att}}$. The weights of the self-attention are obtained by computing the dot-product between the query and all keys and normalizing with softmax. A scaling factor $\sqrt{d^{\text{att}}}$ is used to smooth the distribution:
\begin{align}
    \text{Attention}(Q, K, V) = \text{softmax}\Big(\frac{QK^T}{\sqrt{d^{\text{att}}}}\Big) V.
\end{align}
To capture information from different representation subspaces, multi-head attention (MHA) is used by multiplying the original queries, keys, and values by different weight matrices:
\begin{align}
    \text{MHA}(Q, K, V) &= \text{Concat}([H_h]_{h=1}^{d^{\text{head}}}) W^{\text{head}}, \\
    \text{where } H_h &= \text{Attention} (QW_h^q, KW_h^k, V_h^v W_h^v),
\end{align}
where $d^{\text{head}}$ is the number of heads, and  $W^{\text{head}} \in \mathbb{R}^{(d^{\text{head}}d^{\text{att}}) \times d^{\text{att}}}$ and $W_h^q, W_h^k, W_h^v \in \mathbb{R}^{d^{\text{att}} \times d^{\text{att}}}$ are learnable parameters.

In general, the speech sequence length can be considerably long, making self-attention computationally difficult. For tasks like speech separation and enhancement, the technique of subsampling is not practical as in speech recognition. Inspired by \cite{Effective-Luong2015,Time-Povey2018}, we adjust the self-attention of the Transformers in the masking network to be performed on a local segment of the speech, because those frames have higher correlation. This time-restricted self-attention for the query at time step $t$ is formalized as:
\begin{align}
    \text{Attention}(Q, K', V') = \text{softmax}\big(\frac{QK'^T}{\sqrt{d^{\text{att}}}}\big) V',
\end{align}
where the corresponding keys and values are $K'=K_{t-l:t+r}$ and $V'=V_{t-l:t+r}$, respectively, with $l$ and $r$ here denoting the left and right context window sizes.

\section{Experiments}
\label{sec:foot}
The proposed methods were evaluated on the same dataset as in \cite{MIMO-Chang2019}, referred to as the spatialized wsj1-2mix dataset, where the number of speakers in an utterance is $J=2$. The multi-channel speech signals were generated\footnote{The spatialization toolkit is available at \url{http://www.merl.com/demos/deep-clustering/spatialize_wsj0-mix.zip}} from the monaural wsj1-2mix speech used in \cite{End2End-Seki2018,End-Chang2019}.  The room impulse responses (RIR) for the spatialization were randomly generated\footnote{The RIR generator script is available online at \url{https://github.com/ehabets/RIR-Generator}}, characterizing the room dimensions, speaker locations, and microphone geometry.  The final spatialized dataset contains two different environment conditions, anechoic and reverberant. In the anechoic condition, the room is assumed to be anechoic and only the delays and decays due to the propagation are considered when generating the signals. In the reverberant condition, reverberation is also considered, with randomly drawn T60s from $\left[0.2, 0.6\right]$~s. In total, the spatialized corpus under each condition contains 98.5 hr, 1.3 hr, and 0.8 hr in training, development, and evaluation sets respectively.

In the single-channel multi-speaker speech recognition task, we used the 1st channel of the training, development, and evaluation set to train, validate, and evaluate our model respectively. The input features are 80-dimensional log mel-filterbank coefficients with pitch features and their delta and delta delta features. In the multi-channel multi-speaker speech recognition task, we also followed \cite{MIMO-Chang2019} in including the WSJ train\_si284 in the training set to improve the performance. The model takes the raw waveform audio signal as input and converts it to its STFT using a 25 ms-long Hann window with stride 10 ms. The spectral feature dimension is $F=257$ due to zero-padding. After the frontend computation, 80-dimensional log filterbank features are extracted for each separated speech signal and global mean-variance normalization is applied, using the statistics of the single-speaker WSJ1 training set. All the multi-channel experiments were performed with $C=2$ channels. However, the model can be extended to an arbitrary number of input channels as described in \cite{Multichannel-Ochiai2017}.

\subsection{Experimental Setup}
\label{ssec:setup}

All the proposed end-to-end multi-speaker speech recognition models are implemented with the ESPnet framework \cite{ESPnet-Watanabe2018} using the Pytorch backend. Some basic parts are the same for all the models.  The interpolation factor $\lambda$ of the loss function in (\ref{eq:loss}) is set to 0.2. The word-level language model \cite{End-Hori2018} used during decoding was trained with the official text data included in the WSJ corpus. The configurations of the RNN-based models are the same as in \cite{End-Chang2019} and \cite{MIMO-Chang2019} for single-channel and multi-channel experiments, respectively.

In the Transformer-based multi-speaker encoder-decoder ASR model, there is a total of 12 layers in the encoder and 6 layers in the decoder as in \cite{Improving-Nakatani2019}. Before the Transformer encoder, the log mel-filterbank features are encoded by two CNN blocks. The CNN layers have a kernel size of $3 \times 3$ and the number of feature maps is 64 in the first block and 128 in the second block. For the single-channel multi-speaker model in Sec.~\ref{ssec:monaural-speech}, $\text{Encoder}_{\text{Mix}}$ is the same as the CNN embedding layer, and $\text{Encoder}_{\text{SD}}$ and $\text{Encoder}_{\text{Rec}}$ contain $4$ and $8$ Transformer layers, respectively. For all the tasks, the configuration of each encoder-decoder layer is $d^{\text{att}}=256$, $d^{\text{ff}}=2048$, $d^{\text{head}}=4$. The masking network in the frontend has 3 layers similar to the encoder-decoder layer except $d^{\text{ff}}=768$. The training stage of Transformer runs with the Adam optimizer and Noam learning rate decay as in \cite{Attention-Vaswani2017}. Note that the backend ASR module is currently initialized with a pretrained model from the ESPnet recipe of WSJ corpus and kept frozen for the first 15 epochs, for training stability.

\vspace{-.1cm}
\subsection{Performance in Anechoic Condition}
\label{ssec:exp1}

We first provide in Table \ref{tab:asr_rslt_monaural} the performance in anechoic condition of the single-channel multi-speaker end-to-end ASR models trained and evaluated on the original single-channel wsj1-2mix corpus used in \cite{End-Hori2018,End-Chang2019}. All the layers are randomly initialized. The result shows that using the Transformer model leads to a $40.9\%$ relative word error rate (WER) improvement on the evaluation set, decreasing from $20.43\%$ to $12.08\%$ compared with the RNN-based model in \cite{End-Chang2019}.

The multi-channel multi-speaker speech recognition performance is shown in Table \ref{tab:asr_rslt_anechoic} using the spatialized anechoic wsj1-2mix dataset. The baseline multi-channel system is the RNN-based model from our previous study \cite{MIMO-Chang2019}. Before we move to the fully Transformer-based MIMO-Speech model, we first replace the RNNs with Transformers in the backend ASR only. We see that using Transformers for the ASR backend can achieve $20.5\%$ relative improvement against the RNN-based model in anechoic conditions.

We then also apply Transformers in the masking network of the frontend. Considering the feasibility of computing, in this preliminary study, the left and right context window sizes of the self-attention are set to $l=14$ and $r=15$. 
The parameters of the frontend are randomly initialized. Compared with using a Transformer-based model only for the backend, the fully Transformer-based model leads to a further improvement, achieving a WER of $6.41\%$. Compared against the whole sequence information available in the RNN-based model, such a small context window greatly limits the power of our model but shows its potential. Overall, the proposed fully Transformer-based model achieves a $25.6\%$ relative WER improvement against the RNN-based model in the multi-channel case. We also see that the multi-channel system is better than the single-channel system, thanks to the availability of spatial information.

\begin{table}[]
    \vspace{-0.3cm}
    \small
    \centering
    \caption{Performance in terms of average WER [\%] on the \textbf{single-channel anechoic} wsj1-2mix corpus.}
    \vspace{0.1cm}
    \label{tab:asr_rslt_monaural}
    \begin{tabular}{lcc}
    \toprule
    Model & dev & eval \\
    \midrule
    RNN-based 1-channel Model \cite{End-Chang2019} & 24.90 & 20.43 \\
    Transformer-based 1-channel Model & \textbf{17.11} & \textbf{12.08} \\
    \bottomrule
    \end{tabular}
    \vspace{-0.4cm}
    \normalsize
\end{table}
\begin{table}[]
    \small
    \centering
    \caption{Performance in terms of average WER [\%] on the spatialized \textbf{two-channel anechoic} wsj1-2mix corpus.}\vspace{0.1cm}
    \label{tab:asr_rslt_anechoic}
    \begin{tabular}{lcc}
    \toprule
    Model & dev & eval \\
    \midrule
    RNN-based MIMO-Speech \cite{MIMO-Chang2019} & 13.54 & 8.62 \\
    \midrule
    \, + Transformer backend & \textbf{10.73} & 6.85 \\
    \, \, \, ++ Transformer frontend & 11.75 & \textbf{6.41} \\
    \bottomrule
    \end{tabular}
    \vspace{-0.2cm}
    \normalsize
    \vspace{-0.2cm}
\end{table}

\subsection{Performance in Reverberant Condition}
\label{ssec:exp-wpe}

Even though our model can perform very well in anechoic condition, such ideal environments are rarely encountered in practice.
It is thus crucial to investigate whether the model can be applied in more realistic environments. In this subsection, we describe preliminary efforts to process the reverberated signal.

We first used a straightforward multi-conditioned training by adding reverberated utterances into the training set. The results of multi-speaker speech recognition on the multi-channel reverberant datasets are shown in Table \ref{tab:asr_rslt_reverb}. It can be observed that only using the Transformers for the backend is $6.6\%$ better than the RNN-based model. In addition, the fully Transformer-based model achieves  $13.2\%$ relative WER improvement on the evaluation set, which is consistent with the anechoic case. However, comparing with the numbers for the anechoic condition in Table \ref{tab:asr_rslt_anechoic}, a large performance degradation can be observed.

To alleviate this, we turned to an existing external dereverberation method to preprocess the input signals as a simple yet effective solution. Nara-WPE \cite{NaraWPE-Drude2018} is a widely used open source software for blind dereverberation of acoustic signals. The dereverberation is performed on the reverberated speech before it is added to the training dataset with anechoic data. Similarly, the reverberant test set is also preprocessed. Speech recognition performance on the multi-channel reverberant speech after Nara-WPE is shown in Table \ref{tab:asr_rslt_dereverb}. In general, the WERs are dramatically decreased with the dereverberation method. For the RNN-based model, the WER on the evaluation set decreased by $41.1\%$ relative, from $29.99\%$ to $17.67\%$. Similar to the experiments under  other conditions, the model with backend Transformer only is better than the RNN-based baseline model on the reverberant evaluation set by $13.8\%$ relative WER. However, the Transformer-based frontend slightly degraded the performance. This may be due the window size of the attention being too small, as it only covers about $0.3$ s of speech. Note that our systems are not trained through Nara-WPE, which is left for future work.

At last, we show results in the single-channel task with the 1st channel of the reverberated speech after Nara-WPE dereverberation in Table \ref{tab:asr_rslt_monaural_dereverb}. Using the RNN-based model, the WER of the evaluation set is high, at $28.21\%$, which is influenced greatly by the reverberation, even when preprocessing with the dereverberation technique. However, the Transformer-based model can reach a final WER of $16.50\%$, a $41.5\%$ relative reduction, proving that the Transformer-based model is more robust than the RNN-based model.

\begin{table}[]
    \vspace{-0.3cm}
    \small
    \centering
    \caption{Performance in terms of average WER [\%] on the spatialized \textbf{two-channel reverberant} wsj1-2mix corpus.}\vspace{0.1cm}
    \label{tab:asr_rslt_reverb}
    \begin{tabular}{lcc}
    \toprule
    Model & dev & eval \\
    \midrule
    RNN-based MIMO-Speech \cite{MIMO-Chang2019} & 34.98 & 29.99 \\
    \midrule
    \, + Transformer backend & 32.95 & 28.01 \\
    \, \, \, ++ Transformer frontend & \textbf{31.93} & \textbf{26.02} \\
    \bottomrule
    \end{tabular}
    \vspace{-0.15cm}
    \normalsize
\end{table}
\begin{table}[]
    \vspace{-0.1cm}
    \small
    \centering
        \caption{Performance in terms of average WER [\%] on the spatialized \textbf{two-channel reverberant} wsj1-2mix corpus \textbf{after Nara-WPE}.}
        \vspace{0.1cm}
    \label{tab:asr_rslt_dereverb}
    \begin{tabular}{lcc}
    \toprule
    Model & dev & eval \\
    \midrule
    RNN-based MIMO-Speech & 24.45 & 17.67 \\
    \midrule
    \, + Transformer backend & \textbf{19.17} & \textbf{15.24} \\
    \, \, \, ++ Transformer frontend & 20.55 & 15.46 \\
    \bottomrule
    \end{tabular}
    \vspace{-0.2cm}
    \normalsize
\end{table}
\begin{table}[htb]
    \small
        \caption{Performance in terms of average WER [\%] on the \textbf{1st channel} of the spatialized \textbf{reverberant} wsj1-2mix corpus after Nara-WPE.}
        \vspace{0.1cm}
    \label{tab:asr_rslt_monaural_dereverb}
    \centering
    \begin{tabular}{lcc}
    \toprule
    Model & dev & eval \\
    \midrule
    RNN-based 1-channel Model & 31.21 & 28.21 \\
    Transformer-based 1-channel Model & 20.44 & 16.50 \\
    \bottomrule
    \end{tabular}
    \vspace{-0.2cm}
    \normalsize
\end{table}





\section{Conclusion}
\label{sec:conclusion}

In this paper, we applied Transformer models for end-to-end multi-speaker ASR in both the single-channel and multi-channel scenarios, and observed consistent improvements. The RNN-based ASR module is replaced with the Transformers. To alleviate the fatal memory consumption issue when applying Transformers in the frontend with considerably long sequences, we modified the self-attention in the Transformers of the masking network by using a local context window. Furthermore, by incorporating an external dereverberation method, we largely reduced the performance gap between the reverberant condition and the anechoic condition, and hope to further reduce it in the future thanks to tighter integration of the dereverberation within our model. 

\vfill
\pagebreak
\bibliographystyle{IEEEtran}
\bibliography{strings,refs}

\end{document}